\documentclass[showpacs,amsmath,amssymb]{revtex4}

\usepackage{graphicx}
\usepackage{dcolumn}
\usepackage{bm}
\usepackage{epsfig}

\begin{document}

\def\0#1#2{\frac{#1}{#2}}
\def\bct{\begin{center}} \def\ect{\end{center}}
\def\beq{\begin{equation}} \def\eeq{\end{equation}}
\def\bea{\begin{eqnarray}} \def\eea{\end{eqnarray}}
\def\nnu{\nonumber}
\def\n{\noindent} \def\pl{\partial}
\def\g{\gamma}  \def\O{\Omega} \def\e{\varepsilon} \def\o{\omega}
\def\s{\sigma}  \def\b{\beta} \def\p{\psi} \def\r{\rho}
\def\G{\Gamma} \def\S{\Sigma} \def\l{\lambda}

\title{Studying the baryon properties through chiral soliton model at finite temperature and denstity}
\author{Song~Shu}
\affiliation{Department of Physics and Electronic Technology,
Hubei University, Wuhan 430062, China}
\author{Jia-Rong~Li}
\affiliation{Institute of Particle Physics, Hua-Zhong Normal
University, Wuhan 430079, China} \affiliation{Key Laboratory of
Quark {\rm\&} Lepton Physics (Hua-Zhong Normal University),
Ministry of Education, China}
\begin{abstract}
We have studied the chiral soliton model in a thermal vacuum. The
soliton equations are solved at finite temperature and density.
The temperature or density dependent soliton solutions are
presented. The physical properties of baryons are derived from the
soliton solutions at finite temperature and density. The
temperature or density dependent variation of the baryon
properties are discussed.
\end{abstract}  \maketitle

\section{Introduction}
The fundamental theory of the strong interaction is quantum
chromodynamics (QCD). This theory has the essential properties of
asymptotic freedom, hidden (spontaneously broken) chiral symmetry
and confinement. For the high energy process, the perturbative QCD
calculation works well due to the asymptotic freedom. However, in
the low energy nuclear sector, it is a hard journey for the QCD to
walk out the way to calculate the quantities of nuclear physics
because of the nonperturbative features of broken chiral symmetry
and confinement. There are different approaches in this direction,
like lattice QCD, QCD sum rule, chiral perturbation theory and
Dyson-Schwinger equation~\cite{ref1,ref2,ref3,ref4,ref5}. Another
important approach is to use the effective models which have
certain essential features of QCD. In studying the hadron
properties, the bag models which has the essential feature of
confinement have been often used. They are MIT, SLAC and
Friedberg-Lee models which describe the static nucleon properties
quite well~\cite{ref6,ref7}. In these models the hadronic bound
states is determined by the confinement mechanism. There is also
other perspective about the hadronic bound states. It suggests
that there is a separation of roles between the forces responsible
for binding quarks in hadrons and those which give absolute
confinement~\cite{ref8,ref9}. The forces responsible for quark
binding are strong Coulomb-type and related to vacuum condensation
and dynamical chiral symmetry breaking. Thus these authors
suggested using chiral models to study hadron
properties~\cite{ref10}. The chiral model was solved in the
semiclassical or mean-field approximation and the semiclassical
solution was referred to a chiral soliton. In the environment of
vacuum, they have studied the static properties of nucleon and
delta~\cite{ref9}. After that the subsequential work had been made
to study hadron properties in
vacuum~\cite{ref11,ref12,ref13,ref13a}.

In high energy physics, the nuclear matter is studied under
extreme condition. The study of hot and dense medium created in
high energy heavy ion collisions is of great
interest~\cite{ref14,ref15}. The thermal vacuum is very different
to the vacuum at zero temperature and density. The vacuum
condensation would be melted which will result in the chiral
symmetry restoration and deconfinement of the system. So it is a
very interest topic to study the hadron properties in the thermal
vacuum. It could give us more insight into the complex vacuum
structure of QCD. In the present work our purpose is to extend the
study of the chiral soliton model to finite temperature and
density. The chiral soliton and baryon properties will be studied
at finite temperature and density.

At finite temperature and density the chiral soliton model which
is also called linear sigma model has been used extensively to
study chiral restoration in an uniform
system~\cite{ref16,ref17,ref18,ref19,ref20}, while the studis of
chiral soliton and hadron properties at finite temperature and
density in the same model are not so many. In
reference~\cite{ref21}, the chiral solitons in linear sigma model
and NJL model were studied in hot and dense medium by means of
variational projection techniques. In reference~\cite{ref22},
temperature effect to chiral soliton model was introduced by
adding one loop potential to the Lagrangian. Here we will use the
stand finite temperature field theory to introduce temperature and
density effect. In most recent work~\cite{ref23}, the chiral
soliton model has been discussed through the formalism of finite
temperature field theory. Their main purpose is to study the
chiral soliton in chiral phase transition. Our goal is concentrate
on the baryon properties at finite temperature and density, and
our treatment about the baryon mass is quite different with
theirs, so the result is quite different, which will be discussed
later.

The organization of this paper is as follows: in section 2 the
chiral soliton model is introduced. The field equations are solved
at zero temperature and density. The baryon properties in vacuum
are reproduced. In section 3, the chiral soliton equations are
extended to finite temperature and density. The soliton solutions
are discussed at finite temperatures and densities. In section 4,
the baryon properties are derived from the soliton solutions. The
temperature or density dependent variation of these properties are
presented and discussed. The last section is the summary.

\section{Chiral soliton model at zero temperature and density}
We start from the Lagrangian of the chiral soliton model, \beq
{\cal
L}=\bar\psi[i\gamma_\mu\pl^\mu+g(\s+i\gamma_5\vec\tau\cdot\vec\pi)]\psi+\012(\pl_\mu\s\pl^\mu\s+\pl_\mu\vec\pi\pl^\mu\vec\pi)-U(\s,\vec\pi),
 \label{lag} \eeq where\bea
U(\s,\vec\pi)=\0{\l}{4}(\s^2+\vec\pi^2-\nu^2)^2+H\s-\0{m_\pi^4}{4\l}+f_\pi^2
m_\pi^2. \label{potential} \eea $\p$ represents the two flavor
light quark fields $\psi=(u,d)$, $\s$ is the isosinglet scalar
field, and $\vec\pi$ is the isovector pion field
$\vec\pi=(\pi_1,\pi_2,\pi_3)$. $H\s$ is the explicit chiral
symmetry breaking term and $H=f_{\pi}m_\pi^2$, where $f_\pi=93MeV$
is the pion decay constant and $m_\pi=138MeV$ is the pion mass.
The chiral symmetry is explicitly broken in the vacuum and
expectation values of the meson fields are
$\langle\s\rangle=-f_\pi$ and $\langle\vec\pi\rangle=0$. The
constitute quark mass in vacuum is $M_q=gf_\pi$, and the sigma
mass is $m_\s^2=m_\pi^2+2\l f_\pi^2$. The quantity $\nu^2$ can be
expressed as $\nu^2=f_\pi^2-m_\pi^2/\l$. The last two constants in
equation (\ref{potential}) ensure that the vacuum energy is zero.
In our calculation we have followed the choice of the
reference~\cite{ref9} and set the constituent quark mass and the
sigma mass as $M_q=500MeV$ and $m_\s=1200MeV$ which determine the
parameters $g\approx 5.28$ and $\l\approx 82.1$.

From the Lagrangian (\ref{lag}) the field equations could be
derived in the following, \beq
[i\gamma^\mu\pl_\mu+g(\s+i\gamma_5\vec\tau\cdot\vec\pi)]\psi=0,
\label{q} \eeq \beq \pl_\mu\pl^\mu\s-g\bar\psi\psi=-\0{\pl
U(\s,\vec\pi)}{\pl\s},\label{s} \eeq \beq
\pl_\mu\pl^\mu\vec\pi-ig\bar\psi\gamma_5\vec\tau\psi=-\0{\pl
U(\s,\vec\pi)}{\pl\vec\pi}. \label{p} \eeq One could take the mean
field approximation and the ``hedgehog" ansatz which means, \bea
\s(\vec r,t)=\s(r), \ \ \ \ \vec\pi(\vec r,t)=\hat\vec r\pi(r),
\eea \bea \psi(\vec r,t)=e^{-i\e t}\sum_{i=1}^{N}q_i(\vec r),\ \ \
\ \ q(\vec r)=\left(
\begin{array}{c}
u(r)\\
i\vec\s\cdot\hat\vec r v(r)
\end{array}\right)\chi,
\eea where $q_i$ are $N$ identical quarks in the lowest s-wave
level with energy $\e$. $N=3$ is for baryons and $N=2$ for meson.
$\chi$ is the spinor which satisfies the condition \bea
(\vec\s+\vec\tau)\chi=0.\eea $\hat\vec r$ is the radial unit
vector. The meson fields $\s(r)$, $\pi(r)$ and the quark functions
$u(r)$, $v(r)$ are spherical symmetric and satisfy the following
set of coupled nonlinear radial differential equations which could
be obtained from equations (\ref{q})-(\ref{p}), \beq
\0{du(r)}{dr}=-(\e-g\s(r))v(r)-g\pi(r)u(r),\label{u} \eeq \beq
\0{dv(r)}{dr}=-\left(\02r-g\pi(r)\right)v(r)+(\e+g\s(r))u(r),\label{v}
\eeq \beq
\0{d^2\s(r)}{dr^2}+\02r\0{d\s(r)}{dr}+Ng(u^2(r)-v^2(r))=\0{\pl
U}{\pl\s},\label{s0} \eeq \beq
\0{d^2\pi(r)}{dr^2}+\02r\0{d\pi(r)}{dr}-\0{2\pi(r)}{r^2}+2Ngu(r)v(r)=\0{\pl
U}{\pl\pi}. \label{p0} \eeq The quark functions should satisfy the
normalization condition \bea 4\pi\int r^2(u^2(r)+v^2(r))dr=1.
\label{n} \eea The boundary conditions on the quark functions and
meson fields are, \bea v(0)=0, \ \ \0{d\s(0)}{dr}=0,\ \ \pi(0)=0,
\label{b1} \eea \bea u(\infty)=0, \ \ \s(\infty)=-f_\pi, \ \
\pi(\infty)=0.\label{b2} \eea As mentioned above the parameters
have been fixed to $g\approx 5.28$ and $\l\approx 82.1$. The
lowest quark energy eigenvalue is $\e=30.5MeV$~\cite{ref9}.
\begin{figure}[tbh]
\includegraphics[width=210pt,height=150pt]{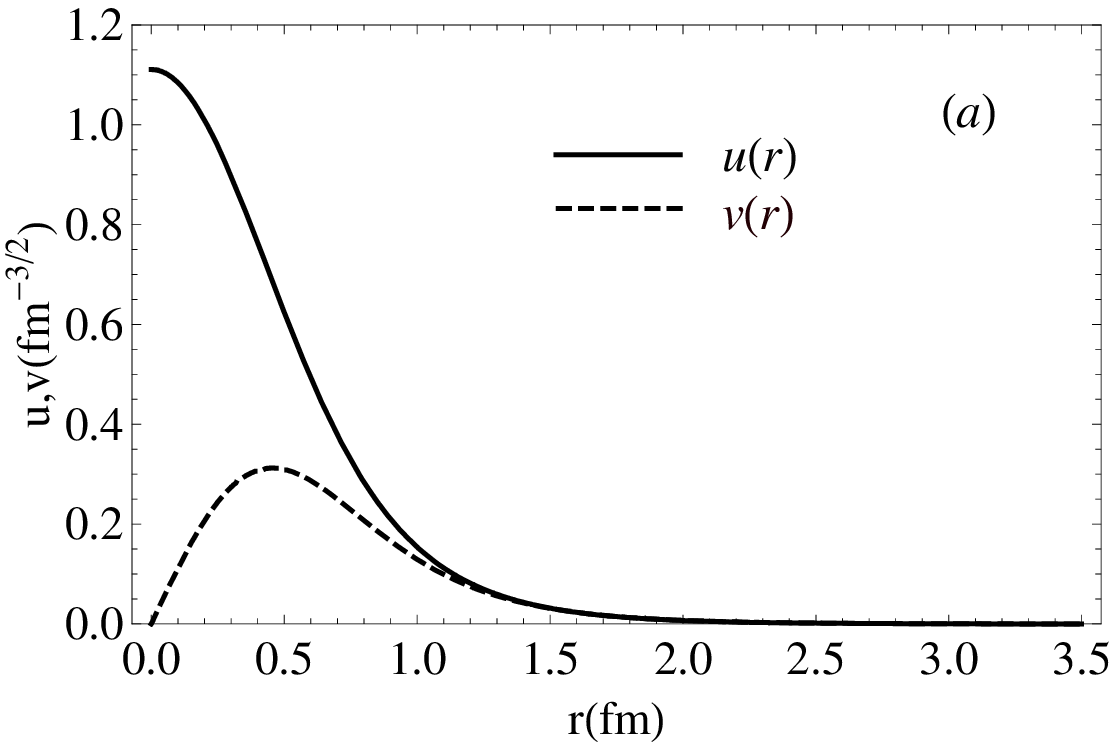}
\hspace{1cm}
\includegraphics[width=210pt,height=150pt]{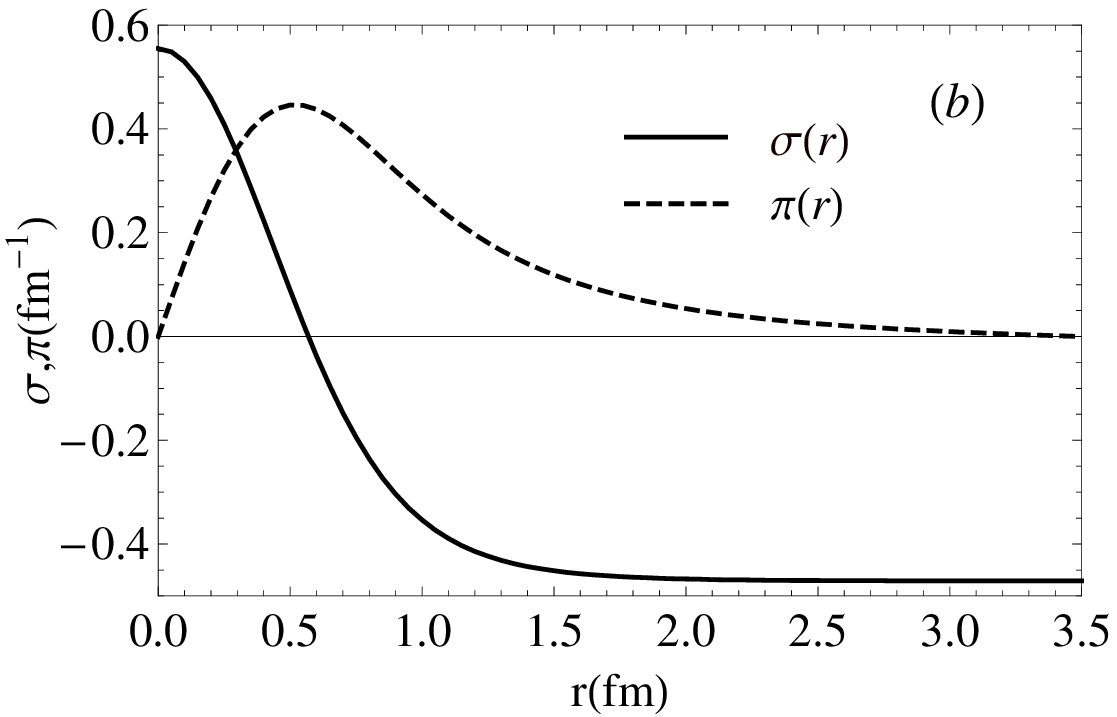}
\caption{The soliton solutions at zero temperature and chemical
potential. (a)The quark fields $u(r)$ and $v(r)$. (b)The meson
fields $\s(r)$ and $\pi(r)$.}\label{f1}
\end{figure}
In our study we are mainly concerned about the properties of
baryons which means $N=3$. The equations (\ref{u})-(\ref{p0})
together with normalization condition (\ref{n}) and boundary
conditions (\ref{b1}) and (\ref{b2}) could be numerically solved
by a standard numerical package which is called
COLSYS~\cite{ref24}. The forms of the fields in a chiral soliton
are shown in Fig.\ref{f1} From these soliton solutions one can
further derived the general properties of a baryon, like the
soliton energy or baryon mass $M_B$, the root mean square charge
radius $r_B$, the magnetic moment $\mu_B$ and the ratio of the
axial to vector coupling constants $g_A/g_V$, which could be
calculated in the following, \beq E=M_B=3\e+4\pi\int
drr^2\left[\012\left(\0{d\s}{dr}\right)^2+\012\left(\0{d\pi}{dr}\right)^2+\0{\pi^2}{r^2}+U(\s,\pi)\right],
\label{mb} \eeq \bea <r_B^2>=4\pi\int_0^\infty(u^2+v^2)r^4dr,
\label{rb} \eea \bea \mu_B=\0{8\pi}3\int_0^\infty r^3uvdr,
\label{mub} \eea \bea \0{g_A}{g_V}=\0{20\pi}3\int_0^\infty
r^2(u^2-\0{v^2}3)dr. \label{gb} \eea In our calculation all these
physical quantities could be well reproduced and the results are:
$M_B=1140MeV$, $r_B=0.717fm$, $\mu_B=0.195efm$ and $g_A/g_V=1.162$
which are in agreement with the results in reference~\cite{ref9}.
All the calculations and results in the above discussions are in
the vacuum which means at zero temperature and density. In the
following sections we will study the solitons and baryon
properties at finite temperature and density.

\section{chiral soliton solutions at finite temperatures and densities}
We consider a baryon embedded in a thermal quark medium. The
vacuum become a thermal vacuum. The equations (\ref{s}) and
(\ref{p}) become, \beq
\pl_\mu\pl^\mu\s-g\bar\psi\psi=g\langle\bar\psi\psi\rangle-\0{\pl
U(\s,\vec\pi)}{\pl\s},\label{s1} \eeq \beq
\pl_\mu\pl^\mu\vec\pi-ig\bar\psi\gamma_5\vec\tau\psi=g\langle
i\bar\psi\gamma_5\vec\tau\psi\rangle-\0{\pl
U(\s,\vec\pi)}{\pl\vec\pi}, \label{p1} \eeq in which, the quark
source terms decompose into two parts: one is the usual source
part of one baryon as $g\bar\psi\psi$ or
$ig\bar\psi\gamma_5\vec\tau\psi$ on the left hand side of the
equation; the other is the thermal quark medium source part as
$g\langle\bar\psi\psi\rangle$ or $g\langle
i\bar\psi\gamma_5\vec\tau\psi\rangle$ on the right hand side of
the equation. The meson field functions $\s(r)$ and $\pi(r)$ are
taken as the classic mean fields in the thermal vacuum. The
thermal quark medium part is also the thermal vacuum average of
the quark source which could be calculated by standard method in
finite temperature theory and the results are, \bea
\langle\bar\psi\psi\rangle &=& -g\s\nu_q\int \0{d^3\bf
p}{(2\pi)^3}\0{1}{E_q}\left(\01{e^{\b(E_q-\mu)}+1}+\01{e^{\b(E_q+\mu)}+1}\right),
\label{s2} \\ \langle i\bar\psi\gamma_5\vec\tau\psi\rangle &=&
-g\vec\pi\nu_q\int \0{d^3\bf
p}{(2\pi)^3}\0{1}{E_q}\left(\01{e^{\b(E_q-\mu)}+1}+\01{e^{\b(E_q+\mu)}+1}\right),
\label{p2} \eea where $\b$ is the inverse temperature $\b=1/T$,
$\mu$ is the quark chemical potential, $\nu_q$ is a degenerate
factor, $\nu_q=2(spin)\times 2(flavor)\times 3(color)$, and
$E_q=\sqrt{{\bf p}^2+m_q^2}$ with an effective quark mass term
$m_q=g\s$. Furthermore we could make the following definitions,
\bea \langle\bar\psi\psi\rangle\equiv -\r_s(T,\mu), \ \ \ \ \
\langle i\bar\psi\gamma_5\vec\tau\psi\rangle\equiv -
\vec\r_{ps}(T,\mu), \eea \bea \0{\pl
U(\s,\vec\pi)}{\pl\s}+g\r_s(T,\mu)\equiv\0{\pl
V_{eff}(T,\mu)}{\pl\s}, \ \ \ \ \ \0{\pl
U(\s,\vec\pi)}{\pl\vec\pi}+g\vec\r_{ps}(T,\mu)\equiv\0{\pl
V_{eff}(T,\mu)}{\pl\vec\pi}, \eea where $\r_s$ and $\vec\r_{ps}$
are the scalar and pseudoscalar densities of quarks and
anti-quarks. $V_{eff}$ is defined as a thermal effective potential
of the quark medium and by its definition the form is \bea
V_{eff}(T,\mu)=U(\s,\vec\pi)-\0{\nu_q}{\b}\int \0{d^3\bf
p}{(2\pi)^3}\left[\ln (1+e^{-\b(E_q-\mu)}) + \ln
(1+e^{-\b(E_q+\mu)})\right]. \label{eff} \eea The effective
potential here is identical to the one loop or mean field
thermodynamical potential of the model which could be also derived
through the partition function in imaginary time formalism of
finite temperature field theory. As we have mentioned at the start
of the section this uniform quark medium is regraded as the
thermal background in which the baryon is embedded. At this time
the meson radial equations at finite temperature and density could
be derived, \beq
\0{d^2\s(r)}{dr^2}+\02r\0{d\s(r)}{dr}+Ng(u^2(r)-v^2(r))=\0{\pl
V_{eff}}{\pl\s}, \label{s3} \eeq \beq
\0{d^2\pi(r)}{dr^2}+\02r\0{d\pi(r)}{dr}-\0{2\pi(r)}{r^2}+2Ngu(r)v(r)=\0{\pl
V_{eff}}{\pl\pi}. \label{p3} \eeq Compared to the radial equations
(\ref{s0}) and (\ref{p0}) at zero temperature and density, it is
convenient to obtain the finite temperature and density equations
by simply replacing the classical potential $U(\s,\vec\pi)$ with
the thermal effective potential $V_{eff}$. The forms of the
equations of quark functions do not change. However one should
notice that those equations are coupled to the meson field
equations at finite temperature and density, which implies that
the baryon is embedded in the thermal quark medium. As a result
the field functions $u(r)$, $v(r)$, $\s(r)$ and $\pi(r)$ are all
functions of temperature and density. The boundary conditions will
not change except for the sigma field. At zero temperature and
density the sigma field asymptotically approaches its vacuum value
$\langle\s\rangle=-f_\pi$ as $r\rightarrow\infty$. While at finite
temperature and density the sigma field asymptotically approaches
the thermal vacuum value $\langle\s\rangle=-\s_v$ which will be
determined by the absolute minimum of the thermal effective
potential (\ref{eff}). At certain temperature and density the
field equations (\ref{u}), (\ref{v}), (\ref{s3}) and (\ref{p3})
together with the thermal effective potential (\ref{eff}) could be
numerically solved by a modified COLSYS.
\begin{figure}[tbh]
\includegraphics[width=210pt,height=150pt]{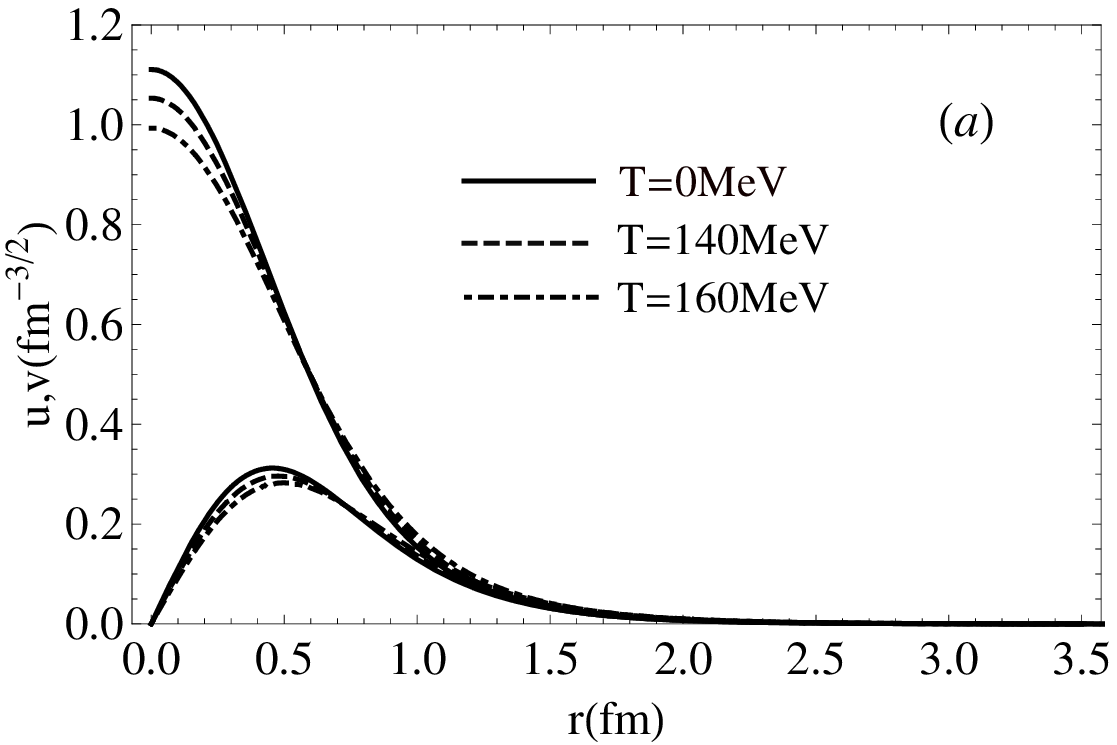}
\hspace{1cm}
\includegraphics[width=210pt,height=150pt]{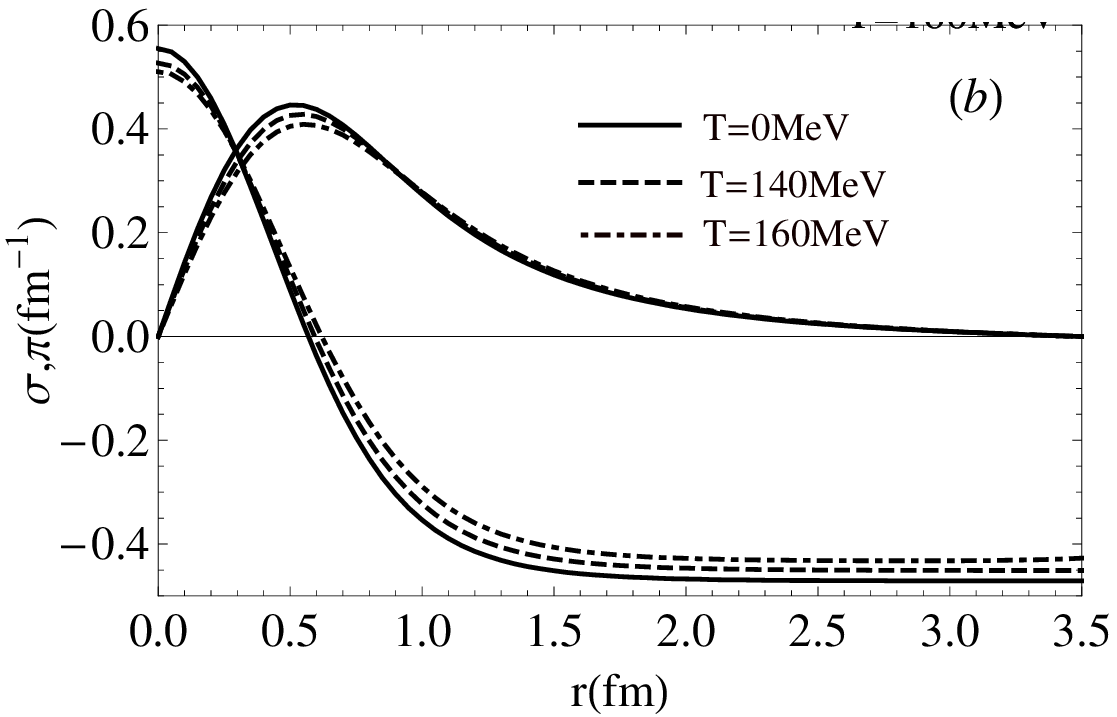}
\caption{At fixed zero chemical potential the soliton solutions
for different temperatures: $T=0MeV$, $T=140MeV$ and $T=160MeV$.
(a)The quark fields $u(r)$ and $v(r)$ for different temperatures.
(b)The meson fields $\s(r)$ and $\pi(r)$ for different
temperatures.}\label{f2}
\end{figure}

In Fig.\ref{f2} we have showed the soliton solutions for different
temperatures at fixed zero density. It could be seen that the
amplitudes of the soliton solutions decrease with the temperature
increasing. At relatively low temperatures $T\lesssim 100MeV$ the
soliton solutions change slowly while at relatively high
temperatures they change more and more quickly. It could be
estimated that when temperature increasing from $0MeV$ to $100MeV$
the soliton amplitudes decrease by $1\%\sim 2\%$; when temperature
increasing from $100MeV$ to $160MeV$ the soliton amplitudes
decrease by $6\%\sim 9\%$.
\begin{figure}[tbh]
\includegraphics[width=210pt,height=150pt]{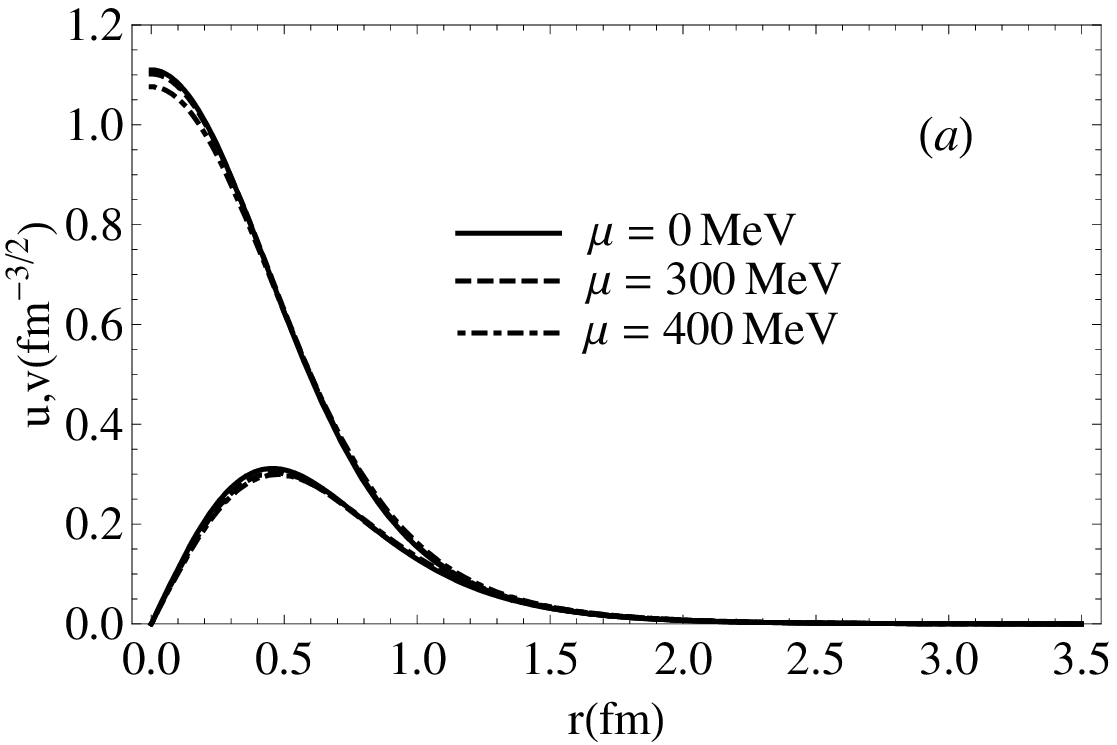}
\hspace{1cm}
\includegraphics[width=210pt,height=150pt]{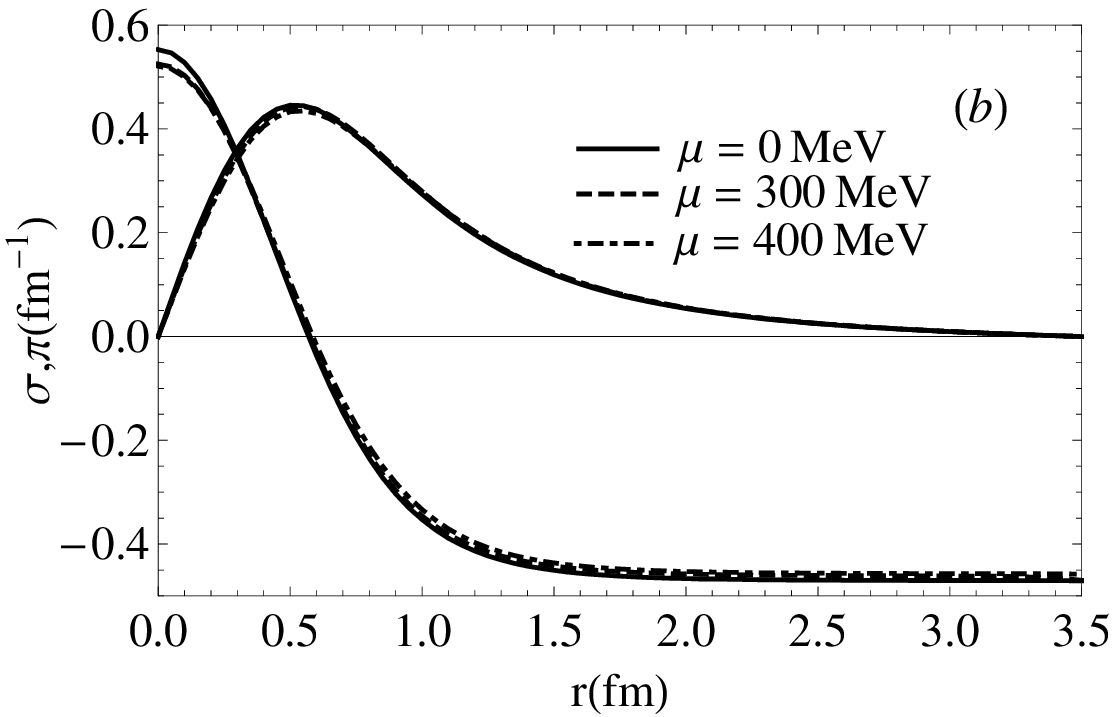}
\caption{At fixed temperature $T=50MeV$ the soliton solutions for
different chemical potentials: $\mu=0MeV$, $\mu=300MeV$ and
$\mu=400MeV$. (a)The quark fields $u(r)$ and $v(r)$ for different
chemical potentials. (b)The meson fields $\s(r)$ and $\pi(r)$ for
different chemical potentials.}\label{f3}
\end{figure}

In Fig.\ref{f3} we have showed the soliton solutions for different
chemical potentials at fixed temperature $T=50MeV$. One can see
that the amplitudes of the solitons decrease slightly with the
chemical potential increasing. When chemical potential increasing
from $0MeV$ to $400MeV$ the soliton amplitudes decrease only by
$2\%\sim 5\%$. When the temperature or density further increases,
there will be chiral restoring phase transition in the system,
which lies out of the scope of this work. In that circumstance the
soliton solutions deserve a thorough investigation in our future
work.

\section{Discussions of baryon properties at finite temperatures and densities}
As the field functions $u(r)$, $v(r)$, $\s(r)$ and $\pi(r)$ at
finite temperatures and densities have been obtained in the above
discussion, in this section we will discuss the baryon properties
at finite temperatures and densities. Substituting the finite
temperature and density field functions $u(r)$, $v(r)$, $\s(r)$
and $\pi(r)$ into the equations (\ref{mb})-(\ref{gb}) one could
calculate the baryon mass $M_B$, the root mean square charge
radius $r_B$, the magnetic moment $\mu_B$ and the ratio of the
axial to vector coupling constants $g_A/g_V$ at different
temperatures and densities.
\begin{table}
\caption{\label{t1}At fixed zero chemical potential the baryon
properties for different temperatures.}
\begin{ruledtabular}
\begin{tabular}{ccccc}
T(MeV)       &  0    & 100   &  140  & 160   \\
\hline
$M_B(MeV)$   & 1140  & 1160  & 1404  & 2096  \\
$r_B(fm)$    & 0.717 & 0.725 & 0.754 & 0.797 \\
$\mu_B(efm)$ & 0.195 & 0.197 & 0.207 & 0.222 \\
$g_A/g_V$    & 1.162 & 1.170 & 1.181 & 1.193 \\

\end{tabular}
\end{ruledtabular}
\end{table}

In Table \ref{t1} we show the baryon properties for the different
temperatures at fixed zero chemical potential. One can see that
all the physical quantities increase with temperature increasing.
In particular the baryon mass increases relatively slowly at
temperature $T\lesssim 100MeV$ while more and more rapidly after
that.
\begin{table}
\caption{\label{t2}At fixed temperature $T=50MeV$ the baryon
properties for different chemical potentials.}
\begin{ruledtabular}
\begin{tabular}{ccccc}
$\mu$(MeV)       &  0    & 200   &  300  & 400   \\
\hline
$M_B(MeV)$       & 1141  & 1164  & 1194  & 1268  \\
$r_B(fm)$        & 0.718 & 0.723 & 0.726 & 0.743 \\
$\mu_B(efm)$     & 0.195 & 0.196 & 0.197 & 0.204 \\
$g_A/g_V$        & 1.163 & 1.175 & 1.185 & 1.189 \\

\end{tabular}
\end{ruledtabular}
\end{table}

In Table \ref{t2} the baryon properties for different chemical
potentials at fixed temperature $T=50MeV$ are presented. It could
be seen that all the physical quantities are also increasing with
chemical potential increasing. The baryon mass increases
moderately and gradually with chemical potential increasing.

Here we make some discussions about the baryon mass. In
reference~\cite{ref23}, the baryon mass is decreasing with
temperature or chemical potential increasing. This is because of
the different schemes in calculating the baryon mass. In our
calculation of the baryon mass as shown in equation (\ref{mb}), we
have included the meson interaction energy $U(\s,\pi)$ while they
have neglected this part and only consider the kinetic energies of
$\s$ and $\pi$. If we neglect the meson interaction energy the
baryon mass is also decreasing in our work. The results about the
baryon mass from these two different schemes are indeed opposite.
However we think that at finite temperature and density the meson
interaction energy should be included as this energy is also a
part of the energy of the baryon. Our results are in agreement
with reference~\cite{ref22} in which they have also taken the
meson interaction energy into account when calculating the baryon
mass. This is an interesting problem of which we will make a study
in detail in a separate work.

\section{summary}
In this paper we have extended the chiral soliton model to finite
temperatures and densities. The chiral soliton equations are
solved and the soliton solutions are discussed at finite
temperatures and densities. The soliton amplitudes decrease with
temperature or density increase. As a result the physical
quantities of the baryon, like the baryon mass, the mean charge
radius, the magnetic moment and the the ratio of the axial to
vector coupling constants, are all increase with temperature or
density increase. The physical quantities change more rapidly at
relatively high temperatures than that in relatively low
temperatures. Compared to the temperature dependent case, the
chemical potential dependent baryon properties change relatively
slowly.

\begin{acknowledgments}
This work was supported in part by the National Natural Science
Foundation of China with No. 10905018 and No. 11275082.
\end{acknowledgments}

\end{document}